\documentclass[prl,twocolumn,floatfix,showpacs,showkeys]{revtex4} 

\usepackage[first,draft]{draftcopy}
\usepackage{latexsym,amssymb,amsmath,amsfonts}
\usepackage{graphicx}
\usepackage{bm}
\newcommand{\beq}{\begin{equation}}
\newcommand{\eeq}{\end{equation}}
\newcommand{\bc}{\begin{center}}
\newcommand{\ec}{\end{center}}
\newcommand{\eeqa}{\end{eqnarray}}
\newcommand{\beqa}{\begin{eqnarray}}
\newcommand{\no}{\noindent}
\newcommand{\pa}{\partial}

\newcommand{\ra}{\rightarrow}

\newcommand{\al}{\alpha}
\newcommand{\be}{\beta}
\newcommand{\ga}{\gamma}

\newcommand{\De}{\Delta}

\newcommand{\la}{\lambda}

\newcommand{\rh}{\rho}
\newcommand{\si}{\sigma}

\newcommand{\ta}{\tau}

\newcommand{\ph}{\phi}

\newcommand{\ed}{\end{document} }

\begin{document}

\title{Negative mass}
\author{Richard T. Hammond}

\email{rhammond@email.unc.edu }
\affiliation{Department of Physics\\
University of North Carolina at Chapel Hill\\
Chapel Hill, North Carolina and\\
Army Research Office\\
Research Triangle Park, North Carolina}

\date{\today}


\begin{abstract}
Some physical aspects of negative mass are examined. Several unusual properties, such as the ability of negative mass to penetrate any armor, are analyzed. Other surprising effects include the bizarre system of negative mass chasing positive pass, naked singularities and the violation of cosmic censorship, wormholes, and quantum mechanical results as well. In addition, a brief look into the implications for strings is given.
\end{abstract}

\maketitle
   
\section{Introduction}
Over the years the concept of negative mass has arisen many times. Sometimes the arguments are simple, if charge can be plus or minus, why not mass? Other times they arise in connection to the negative energy states of the Dirac equation, or the mass reversal of Tiomno.\cite{tiomno},\cite{necessity} Although there is no direct empirical evidence of their existence, many thought experiments have been devised and they violate no fundamental law of physics.

The state of knowledge of the universe, with 90\% of the mass of unknown type and the powerful dark energy increasing our rate of acceleration, pushes us to examine new kinds of matter and energy. In order to judge whether negative mass may exist, or how to find it, it is necessary to know how it behaves.
This article represents an introduction to negative mass, examining its behavior in various physical contexts. 

Negative mass is sometimes dismissed as being non-physical. In a sense this is right since it has never been observed. However, it is also argued that if negative energy states are available, then positive mass would decay away to minus infinity and there would be no observable universe. This is more conjecture than anything, one must look at the particular interactions to judge what negative mass will do. Details concerning these questions may be found in the literature.\cite{borde}

The definition of negative mass adopted here is this: Wherever a mass term $m$ appears, it is replaced by $-m$. Also, by convention of this paper, all letters representing mass $m, \ M$, etc. are positive real numbers. So if we consider Newton's law of motion $F=ma$ we know this is for a positive mass particle. For a negative mass particle, we have $F=-ma$.

 It is interesting to note both the Army and Air Force supported research in General Relativity during the approximate period 1950 to early 1970s. Sir Hermann Bondi was supported by what was then called the Aeronautical Research Lab which was part of the Air Force (today ARL stands for the Army Research Lab), and the Air Force had a documented interest in ``anti-gravity.''\cite{anthology} It was during this period Bondi published his famous article concerning negative mass.\cite{bondi}
 
NASA had a brief program  called The Breakthrough Propulsion Physics Project  from 1996 through 2002. Grants were given to investigate revolutionary new schemes for the propulsion of spacecrafts.\cite{nasa} One of the research programs involved the study of wormholes, and it is known negative mass can keep a wormhole open.\cite{morris}

As a final introductory comment, what follows is by no means meant to be an exhaustive treatise on negative mass. It is meant as an introduction to the subject by using examples from varying subfields within physics.

\section{newtonian physics}

The first question one might ask is, what would an object of negative mass do in the gravitational field of the earth. It is fun to try and answer these kinds of questions by  intuition before figuring them out. For example, you might rely on the principle of equivalence and say a negative mass apple would fall down, just like a positive mass apple.  Newton's law of gravitation for positive mass $m$ is ${\bm F}=-mMG/r^2 \hat{\bm  r}$ where $M$ is the mass of the earth, say. Now suppose we consider an object of negative mass in the positive mass field of the earth. The convention used here is that the force is given by
 ${\bm F}=mMG/r^2 \hat{\bm  r}$ where $m>0$.
 
 So we see the force on our negative mass apple is up, away from the earth. Now, to compute the acceleration, we use Newton's law of motion, which, for a negative mass object in light of the convention is ${\bm F}=-m{\bm a}$. Thus the acceleration is in the opposite direction as the force and so, the negative mass apple falls down. By the same reasoning you may show on a negative mass planet, both negative mass and positive mass fall up.
 
 It is interesting to consider a negative mass object near a positive mass of equal magnitude.  As we saw above, the negative mass is attracted to the positive mass, but the positive mass is repelled by the negative mass, so we find, {\it negative mass chases positive mass}. Thus if a negative mass particle is created near a positive mass particle, the pair will accelerate, approaching the speed of light. One may check both energy and momentum are conserved for such an unusual ``particle.'' As mentioned, Bondi did the general relativistic calculation.\cite{bondi}

 Suppose these two particles were very far apart, where we assume the potential energy to be zero.  Suppose also the negative mass particle is fixed and the positive mass particle is pushed to a distance $a$ of the negative mass. The work it takes is $m^2G/a$, which is the potential energy, and which is also obtained by our simple rule of letting $m\ra-m$. Thus to create such a pair would take the energy $m^2G/a$ and one can speculate on models of real particles, but so far we have restricted ourselves to gravitational interactions.

  Our notions about bound states and total energy should be checked. We say that a positive mass attracts negative mass, so the a negative mass planet, $-m$, could orbit the sun of mass $M$. In this case the potential energy is $mMG/r$ which is positive, but kinetic energy is $-mv^2/2$. One may show the virial theorem, kinetic energy is equal to minus one half the gravitational potential energy, holds, but in this case the total energy is positive.
  
\subsection{antigravity?}

	It has long been speculated that negative mass may be used to shield gravity. The value of such a thing is self evident, but is it possible? Suppose we had a thin negative mass carpet of density $-\si$ gm cm$^{-2}$. We already noted that negative mass falls toward the earth, and so will this carpet. The gravitational field, $\bm g$, outside an isolated carpet is $\bm g= 2G\si\hat{\bm r}$ where as always $\hat{\bm r}$ points away from the mass. This is valid for an infinite carpet, or approximately true close to the center of the finite carpet.
	
	Now consider the gravitational field above the carpet which is lying on the ground, fixed in place (somehow). The field is

\beq
{\bm g}=-\frac{MG}{R^2}\hat{\bm r}+2G\si \hat{\bm r}
\eeq
where $M$ and $R$ are the mass and radius of the earth.
This vanishes if $\si=M/2R^2$, which is about $7\times 10^9$ gm cm$^{-2}$. Thus we could float an object of any mass, but it would take the (negative) mass of a small asteroid to make one square meter, so we may rule out this form of antigravity.
  
 \section{Non-gravitational interactions}
 
 One of the most amazing properties of negative mass has been sometimes referred to as its armor piercing ability. The stronger the armor, the more quickly the negative mass will disintegrate it. To see this, suppose a negative mass object is traveling to the right and it hits a wall. The wall, through electromagnetic forces and quantum mechanics, exerts a force to left, just as it would for a positive mass. But when we apply Newton's law of motion to the negative mass object, ${\bm F}=-m{\bm a}$, we see that it accelerates to the right, opposite the direction of the force acting on it. Now the electronic orbitals are squeezed even more,   exerting an ever increasing  force on the object. And so it will blast though the wall, accelerating all the while (for the sake of argument we assume the negative mass is ``stronger'' than the wall, say negative mass steel against concrete).
 
 This would explain why negative mass on earth has  not been found.  First, it would fall to the earth. Then it would receive a mighty boost as it drills it way through the earth, being flung out far above the escape velocity on the other side.
 
 On the other hand, if negative mass is found one would have to be careful in handling it. For example, if you are playing catch with a negative mass ball and it is coming toward you, you must reach behind the ball and pull it towards you to slow it down. Throwing it would take new found skills as well.
 
 There are a host of additional interesting problems one may work out. Consider a positive mass particle with plus charge and an equal but negative mass negatively charged particle. This time, ignoring the gravitational force, we find the positive charge chases the negative charge.

Another interesting idea is the toy model of the photon. Consider a particle of positive mass $m$ and plus charge $e$ and an equal but negative mass particle of minus charge $-e$. Let us suppose they are created a distance $L$ apart and ask, what would happen. First we see the net mass is zero so the total energy $U$ is

\beq
U=\frac{Gm^2-e^2}{L}
 \eeq
 and is positive if $m>e/\sqrt{G}$. In terms of the Planck mass $L_P=\sqrt{G\hbar/c^3}$, this condition is $m>2\sqrt{\al}m_P$ where $\al=e^2/\hbar c$ is the fine structure constant and the Planck mass is $m_P=c^2L_P/2G$. These are Newtonian considerations but the actual numerical results are not important, they show there is a continuum of positive (and negative) energy solutions. In the spirit of negative mass, let us look at the concept of negative energy photons, i. e., the case $m<2\sqrt{\al}m_P$. Let us further suppose the gravitational force is negligible compared to the electromagnetic force.
 
 Suppose we look at the equation of motion of the negative mass particle. The equation of motion is
 
 \beq\label{eqm}
 \frac{dv^\si}{d\ta}=\frac{e}{mc}F^{\si\mu}v_\mu
 \eeq
where $F^{\mu\nu}$ is the electromagnetic field tensor.\cite{jackson} Note the minus signs in $e/m$ cancel, so we have the same equation of motion for the positive mass particle, as expected.
Assuming the motion is along the $x$ direction we call $v^1=dx/d\ta$ where $\ta$ is the proper time, and $v^0=cdt /d\ta$. 

The equations of motion (\ref{eqm}) become

\beqa
 \frac{dv^0}{d\ta}=\frac{e}{mc}F^{01}v_1\\ \nonumber
 \frac{dv^1}{d\ta}=\frac{e}{mc}F^{10}v_0
 \eeqa
 and the solutions are

 \beqa
 v^0=c\cosh \Omega \ta\\ \nonumber
 v^1=c\sinh \Omega\tau
 \eeqa
 
 \no where $\Omega= eE/mc$ and the magnitude of the 
 constant electric field is $E=e/L^2$. This is the well-known case of hyperbolic motion. The results above are given in the rest frame of the particle with proper time $\ta$. The relation to lab frame time $t$ is obtained from the definition of $v^0$ given above, so that
 
 \beq
 \frac{dt}{d\ta}=\cosh\Omega\ta
 \eeq
 so
 \beq
 t=\frac1\Omega \sinh\Omega\tau
 .\eeq
 Using this we may find the velocity as measured in the lab $v_{\mbox{lab}}=dr/dt$, which becomes
 
 \beq
 v_{\mbox{lab}}=\frac{c\Omega t}{\sqrt{1+\Omega^2 t^2}}
 .\eeq
 The time scale controlling the dynamics is $\Omega$ which is
 
 \beq
 \Omega=\frac{e^2}{mcL^2}
 .\eeq
If we take some reasonable value for $L$, say $10^{-15}$ cm, $\Omega$ is of the order of $10^{27}$ sec$^{-1}$, which shows in an immeasurably short time the object is traveling at essentially the speed of light.

 \section{quantum mechanics}

 Consider the simple problem of a particle of mass $m$ in a one dimensional box of length $L$ and height $V$. The Schr\"odinger equation for positive mass  reduces to
 
 \beq
 \frac{d^2\psi}{dx^2}+\frac{2m(E-V)}{\hbar^2}\psi=0
 \eeq
 with $V=0$ inside. Suppose the total energy $E$ (positive) is less than $V$.
 Then we know  there are (depending on the parameters) oscillatory solutions inside the well and decaying exponentials outside. We interpret this as a bound particle.

For negative mass we have
 
  \beq
 \frac{d^2\psi}{dx^2}-\frac{2m(E-V)}{\hbar^2}\psi=0
 \eeq
 and the above remarks are reversed. We expect this, because it is telling us the particle is repelled from the well.
 
 It is interesting to consider a square hill (opposite of well, it tends to push  particles away) of height $V$. One may show this can yield oscillatory solutions inside the hill and decaying exponentials outside. This is the  quantum mechanical analog of a small negative mass particle near a large negative mass particle.
 
 These are quite general considerations. Let's look at the hydrogen aMary. Instead of re-deriving everything that is already in textbooks, we will follow along the standard derivation\cite{merzbacher} and note the radial function $u$, for large $r$, is
 
 \beq
 \frac{d^2u}{dr^2}+\frac{2mE}{\hbar^2}u
 .\eeq
From this we observe spatially confined (bound) states occur when $E$ is negative. For a negative mass electron it is just the opposite. Bound states exist, but they have total positive energy. 
 As a suggested problem one may investigate whether negative mass positrons are bound to the proton, making an interesting charged hydrogen-like aMary.
 
 Relativistic quantum mechanics is governed by the Dirac equation,
 
\beq\label{dirac2}
\ga^a D_a\psi+\frac{imc}{\hbar}\psi=0
\eeq
where $D$ represents the covariant derivative describing the interaction. I will not pursue this further except to say the mass reversal $m\ra-m$ mentioned in the introduction, from a local gauge theory point of view, has recently been shown to be associated with the intrinsic spin of the particle.\cite{necessity} 

 \section{general relativity}
 
 One of the most fascinating predictions of general relativity is the black hole. The solution for the metric tensor, $g_{\mu\nu}$, exterior to a spherically symmetric mass is
 
\beqa\label{metric}
ds^2=g_{\mu\nu}dx^\mu ds^\nu\\ \nonumber
\left(1-\frac kr\right)dt^2-d\Omega^2-\frac{dr^2}{1-\frac kr}
\eeqa
where $k$ is a constant of integration and $ds =cd\ta$ and, as before, $\ta$ is the proper time. To make this reduce to Newtonian physics we must set $k=2MG/c^2$.  

This solution has embarrassed and fascinated us since Schwarzchild derived it. The center of interest is the apparent singularity at $r= r_h$. It was embarrassing in the early days of general relativity because singularities indicate the breakdown of physics. But if, for example, we put $M$ as the sun's mass, $r_h$, which is called the event horizon, is about 3 km. But remember, this is valid only exterior to the mass, so the sun would have to be squeezed into a volume of radius less than 3 km. Stars were known to be much bigger, so this situation was not widely studied.

When it was realized stars could collapse there was renewed interest. As an example of the physical significance in general relativity, the $r,\theta,\ph,  t$  are called coordinate values and are  the same as those very far away. For example if $dr =d\theta=d\ph=0$, then (\ref{metric}) gives

\beq\label{red}
d\tau=\sqrt{1-\frac {2MG}{c^2r}}dt
\eeq

\no  which is telling us time, as measured by an observer near the sun, $d\ta$, runs slower than time as measured by someone far away, who measures $dt$.
Einstein suggested this as a test of his theory and now is considered one of the basic tests that proves his theory.

One may ask, what happens as $r\ra r_h$. It seems time stands still, so this is also called the surface of infinite redshift. All this is discussed extensively in textbooks, and perhaps the most well-known result is, once an observer falls within the event horizon, there is no escape.
To this day this sparks controversy, particularly when one asks about where the information went that fell into the black hole.

The result of all this is, the singularity at the origin ($r=0$) where physics does indeed breakdown, is hidden from our universe. If a singularity is not hidden from sight by the event horizon, then it is called a naked singularity. I should add the apparent singularity at $r=r_h$ is not a singularity, but a manifestation of the coordinate system. For example, the curvature tensor, which determines whether space is curved, is well defined there, although it is singular at the origin.\cite{weinberg}

Finally let us consider Mary and Homer, twins who are far from the black hole. Homer stays home, but Mary decides to explore the black hole and lets herself fall in. What does Homer see? Homer calculates it takes an infinite amount of time for Mary to reach $r_h$, but as far as Mary is concerned she falls right through the event horizon, after which, she can no longer communicate to Homer.\cite{rescue}

Now we may consider the Schwarzchild solution  for negative mass. We saw above that negative mass repels negative mass, so collapse is not possible. But there has been interest in the meaning of the Schwarzchild 
solution for negative mass particles.\cite{on} In addition, there is another possibility: A collection of charged negative mass particles will be self attractive and could in principle collapse! In this case the metric is given by the Reissner-Nordstr\"om solution. Far away the gravitational effects of the charge fall off faster than the mass term, so in this limit it looks like the usual Schwarzchild solution.

In any case for negative mass (\ref{red}) becomes

\beq\label{redneg}
d\tau=\sqrt{1+\frac {2MG}{c^2r}}dt
\eeq
and we see there is no coordinate singularity. In other words, there is no event horizon. There is no black hole.

In the often ironic development in science, it is interesting to see the reaction to the concept of a naked singularity. At first, many did not believe black holes existed. Now, such  objects are observed at the center of galaxies and are taken for granted, but naked singularities are anathema. They expose a breakdown in physics (the singularity) to the outside world that is so unacceptable, to some, that the  cosmic censorship hypothesis has been formulated, which tells us all singularities are hidden. For some of the controversy one may consult a few recent papers.\cite{torri}\cite{unstable}\cite{stable}

Now, with a negative mass Schwarzchild solution (it is tempting to say negative mass black hole, but it is not a black hole if the mass is negative), one may calculate how long Homer sees it takes Mary to reach $r_h$, and it is a finite number, no longer infinity as for the positive mass. In fact, Mary may dip under $r_h$ and zoom back out, as long as the tidal forces were not too big.

If such objects do exist they are nothing like black holes. In fact they are more like cosmic particle accelerators. For example, if a like charge mass were created near this object it would be expelled, and if it were created near $r=0$ it could acquire enormous energy.

For positive mass let us write (\ref{red})
as

\beq\label{time}
\De t=\frac{\De\tau}{\sqrt{1-\frac {2MG}{c^2r}}}
.\eeq

This formula describes particles at rest exterior to an object of mass $M$.
An example would be the atoms on the surface of the sun (neglecting their brownian motion).  Those atoms measure proper time $\tau$. Far away, say the earth, we measure $t$, so if an interval $\De\tau$ goes by on the sun, we measure a time $\De t$ which is longer than $\De\tau$. 
 This is the well-known redshift effect. As $r\ra r_h$, $\De t$ goes to infinity.

  For a negative mass star we have

\beq\label{time}
\De t=\frac{\De\tau}{\sqrt{\left(1+\frac {2MG}{c^2r}\right)}}
\eeq
which shows the light emanating from the star is blue shifted. As $r\ra r_h$, $\De t=\De\ta/\sqrt{2}$ but the singularity at the origin is a point of infinite blue shift.
 
Perhaps the most intriguing  aspect of black holes is the wormhole, also known as the Einstein-Rosen bridge. It turns out the coordinates used to display the solution (\ref{metric}) cover only half of spacetime, there is another distinct region that is asymptotically flat, called a white hole, that is connected by the wormhole. This other region of space could be anywhere in the universe and so, there is the fascinating theoretical possibility of dropping through a wormhole and exiting somewhere quite distant, and do this in much less time than it would take a light ray, not taking the wormhole, to reach there. So, can we reach Andromeda in an hour?
Maybe. Details of the wormhole have been worked out in the Am. J. Phys. article cited.\cite{morris}

One of the problems is that the wormhole is a not a static structure, it is violent and pinches off in a short time. Thus, means must be found to keep it open. In order to accomplish this some kind of exotic matter must be used, and negative mass is a possibility.
 
\subsection{Cosmology}

One of the most amazing properties of Einstein's equations is they can be applied to the universe as a whole and describe its properties. Einstein was the first to do this but it was before the systematic redshift of distant galaxies was discovered by Vesto Slipher and Edwin Hubble, which show the universe is expanding.
  
  Einstein's theory of general relativity reads
  
  \beq\label{ein}
  G^{\mu\nu}=\frac{8\pi G}{c^4}T^{\mu\nu}
 \eeq

 \no where $G^{\mu\nu}$, the Einstein tensor, is a second order nonlinear differential operator of the metric tensor $g_{\mu\nu}$, and the energy momentum tensor on the right hand side represent matter. In the Schwarzchild solution, $T^{\mu\nu}$ is zero, but in cosmological scenarios it is not.

 This is not the place to go into great detail, but we may mention the simple dust model of the universe in which $T^{\mu\nu}=\rho v^\mu v^\nu$ which assumes matter, on a cosmic average, is described by a uniform mass density $\rho$ which exerts no pressure.

 Einstein tried to find a static solution, including the pressure terms, but came to the conclusion there was no such solution to (\ref{ein}), and unable to divine future observations, modified these equations by adding $\la g_{\mu\nu}$, where $\la$ is known as the cosmological constant. In essence the cosmological constant balanced the gravitational self attraction, or in other words, the cosmological constant provided a repulsive force against gravity.

 In the 1920s Alexander Friedmann discovered a time dependent solution and the cosmological constant was no longer needed.
 In this case there are two unknowns, the density $\rho$ and the distance parameter or what is called the radius of the universe $a(t)$. In addition there are three kinds of possible geometries, flat space (but not flat spacetime), space of positive curvature, and space of negative curvature. These are characterized by $k=0, 1, -1$ respectively.
 
 Einstein's equations (\ref{ein}) yield, for positive mass,

 \beq\label{11}
 8\pi G\rh=3\left(\frac{a'}{a}\right)^2+\frac{3kc^2}{a^2}
 \eeq
 
 \beq\label{22}
 4\pi G\rh=-3\frac{a''}{a}
 \eeq
 where $a'=da/dt$.
 
 The solutions are given in any text on general relativity (see \cite{weinberg} for example) and it gives the following physical picture. The universe began with a big bang and expanded to its present day. However, since the matter in the universe is attractive, trying to pull itself together, the rate of expansion was slowing down. To describe this we have the deceleration $q$ is defined as $q=-a''a/a'^2$, which, due to the built in minus sign, is positive if the rate of expansion is slowing.
 
 Until recently this is exactly what was observed, but due to the development of  CCD cameras, and a better understanding of supernovae, Riess\cite{riess} and Perlmutter\cite{perlmutter} turned this picture upside down. In his Nobel Lecture Perlmutter said, ``Apparently we don't live in
a Universe that is currently slowing in its expansion, but rather a Universe
with one of the more interesting histories ... Its
expansion rate used to be slowing, but has been speeding up for the last
half of its history, and presumably could speed up forever."\cite{perlmutternl} Riess put it this way, ``But what I initially measured and wrote in my lab notebook
in the fall of 1997 was stunning! The only way to match the change in the
expansion rate I was seeing was to allow the universe to have a ÒnegativeÓ
mass. In other words, up-ending the equation, the Universe wasn't
decelerating
at all - it was accelerating...!''\cite{riessnl}

 So,  let us consider a negative mass universe,
  
  \beq\label{1}
 -8\pi G\rh=3\left(\frac{a'}{a}\right)^2+\frac{3kc^2}{a^2}
 \eeq
 
 \beq\label{2}
 -4\pi G\rh=-3\frac{a''}{a}
. \eeq
 
 Looking at (\ref{1}) we see that $k$ must be negative. The deceleration parameter, from (\ref{2}), is negative, indicating a universe that  universe has always been  increasing. Today the ``dark energy'' is used to describe the mysterious force pulling our universe apart at an ever increasing rate. There have been many conjectures, but no one is sure what dark energy is.\cite{hamentropy}

 Einstein's theory not only applies to cosmology as a whole but also to astrophysical problems on the cosmic scale. For example it is well know that since gravity curves space, light rays passing around a galaxy can be curved, giving the effect called lensing. Lensing effects due to negative mass have been considered in the literature.\cite{torres}
 \section{strings}
 
 String theory, by which is really meant superstring theory, is a well studied branch of research. Whether or not it is at all a realistic description of nature is yet to be determined. One of the bizarre results of string theory is that it only makes sense in 10 or 11 dimensions. This result is due to the quantization conditions, and there is no fundamental problem with having strings in three dimensional space (four dimensional spacetime). Cosmic strings are one example\cite{vilenkin} and general relativity is another.\cite{shs}
  
The physics of a non-relativistic string is well-known. The wave equation 
for the displacement $y$ is
\beq
\frac{\pa ^2y}{\pa x^2}-\frac\mu T \frac{\pa^2y}{\pa t^2}
\eeq  
where $\mu$ is the mass per unit length and $T$ is the tension in the string.
This will admit wave solutions as long as $\mu$ and $T$ have the same sign. Thus we consider negative mass strings with negative tension.

 Similar ideas hold for the relativistic string, but before we tackle that, let us consider the action
 of a relativistic point particle of positive mass:
 
 \beq\label{point}
 I=-mc^2\int d\tau
 ,\eeq
 which sweeps out a line in spacetime.
 
 Now let us consider a string with coordinates $\xi^\mu$ with $\mu=0,1$ which exists in what is called a target space, in this case our usual Minkowski spacetime with metric $\eta_{\mu\nu}$. 
 In this case the object sweeps out a world-sheet in spacetime.
 The induced metric on the world-sheet is
 
 \beq\label{gam}
 \gamma_{\al\be}=
  \eta_{\mu\nu}\frac{\pa x^\mu}{\pa \xi^\al} 
  \frac{\pa x^\nu}{\pa \xi^\be}
 \eeq
 where $\mu$ and $\nu$ run from 0 to 3 (target space) while $\al$ and $\be$ run over the two string coordinates and, in $(\ref{gam})$, $x^\mu=x^\mu(\xi^0,\xi^1)$.
 Let us call $\xi^0=\ta$ and $\xi^1=\si$. The action of the string may be written as
 
 \beq\label{string}
 I=-T\int \int d\ta d\si\sqrt{-\ga}
 \eeq
 where $\ga=\mbox{det}(\ga_{\al\be})$.
 
 If we assume momentarily $\sqrt{-\ga}$=1, as it would be in a flat two dimensional space, and integrate along the string such that $\int_0^Ld \si=L$ then, with $T=mc^2/L$, we see (\ref{string}) reduce to (\ref{point}).  For negative mass, everything follows with $m$ replaced by $-m$ and a negative tension.
 
The concept of negative tension makes sense for a negative mass string. We saw before that negative mass particles repel each other, so this naturally extends to the concept of a negative mass string expanding, as negative tension would indicate.

Cosmic strings are relativistic strings but, unlike those of string theory, are of cosmic proportion. They have a fascinating property and effect space much differently than point mass matter. Perhaps the most well-known result is exterior to a straight string. The spacetime is flat. Thus, unlike ordinary matter, straight strings do not curve space, yet they are detectable! The concomitant peculiarity is that if go you around the string at fixed distance $r$ exactly once, the distance is $C=(2\pi-\De)r$ where
$\De=8\pi GT/c^4$ is called the deficit angle.

Such an effect may be seen from the apparent direction of distance known stars, the rays of which pass near the string. For a negative mass (negative tension)  $\De$ becomes the surplus angle and such effects may someday be measured.

\section{summary}
The concept of negative mass and some physics of negative mass particles was examined. Some of the unusual effects examined were, negative mass chasing positive mass, and positive charge chasing negative mass. Other effects described were the armor piercing property of negative mass, quantum mechanical effects, and the concept of a mass reversed black hole, which we saw was not  black hole. Finally some introductory remarks about negative mass strings were made.

\ed

http://www.nobelprize.org/nobel_prizes/physics/laureates/2011/riess_lecture.
pdf"
But what I initially measured and wrote in my lab notebook
in the fall of 1997 was stunning! The only way to match the change in the
expansion rate I was seeing was to allow the universe to have a ÒnegativeÓ
mass. In other words, up-ending the equation, the Universe wasnÕt
decelerating
at all - it was accelerating (Fig. 4)!

The initial results indicated the dominating presence of negative mass
accelerating the Universe. Since there is no such thing as negative mass I
introduced the next best thing, the famous cosmological constant to the fit
in
desperation (in the same spirit Einstein had introduced it long ago) and
immediately found that its dominating presence (i.e., a nonzero
vacuum energy with negative pressure causing repulsive gravity) could
explain the apparent acceleration I was seeing

 Riess

-----Original Message-----
From: Hammond, Richard T CIV (US) 
Sent: Wednesday, July 31, 2013 1:13 PM
To: 'rhammond@email.unc.edu'
Subject: quote (UNCLASSIFIED)

Classification: UNCLASSIFIED
Caveats: NONE

Nobel Lecture, December 8, 2011
by

\